\documentclass[aps,superscriptaddress,twocolumn,floatfix,pra,nofootinbib,a4paper]{revtex4-1}
\usepackage{times}
\usepackage{amsfonts}
\usepackage{amsmath}
\usepackage{amssymb}
\usepackage{amsthm}
\usepackage{color}
\usepackage{multirow}
\usepackage[normalem]{ulem}
\usepackage{natbib}
\usepackage[T1]{fontenc}
\usepackage{float}
\usepackage{mathtools}
\usepackage{bm}
\usepackage{graphicx}
\usepackage{xcolor}
	\definecolor{carmine}{RGB}{150,0,24}
\usepackage[colorlinks=true,linkcolor=blue,citecolor=carmine,urlcolor=blue]{hyperref}

\DeclareMathOperator{\Tr}{tr}

\newcommand{\ket}[1]{|#1\rangle}

\newcommand{\ketbra}[2]{|#1\rangle\langle#2|}
\newcommand{\expect}[1]{\langle#1\rangle}

\begin{document}

%%%%%%%%%%%%%%%%%%%%%%%%%%%%%%%%%%%%%%%%%%%%%%%%%%%%%%%%%%%%%%%%%%%

\title{Full network nonlocality}

\author{Alejandro Pozas-Kerstjens}
\affiliation{Departamento de An\'alisis Matem\'atico, Universidad Complutense de Madrid, 28040 Madrid, Spain}
\affiliation{Instituto de Ciencias Matem\'aticas (CSIC-UAM-UC3M-UCM), Madrid, Spain}
\author{Nicolas Gisin}
\affiliation{Group of Applied Physics, University of Geneva, 1211 Geneva 4, Switzerland}
\affiliation{Schaffhausen Institute of Technology -- SIT, Geneva, Switzerland}
\author{Armin Tavakoli}
\affiliation{Institute for Quantum Optics and Quantum Information -- IQOQI Vienna, Austrian Academy of Sciences, Boltzmanngasse 3, 1090 Vienna, Austria}
\affiliation{Institute for Atomic and Subatomic Physics, Vienna University of Technology, 1020 Vienna, Austria}

\begin{abstract}
Networks have advanced the study of nonlocality beyond Bell's theorem. Here, we introduce the concept of full network nonlocality, which describes correlations that necessitate all links in a network to distribute nonlocal resources. Showcasing that this notion is stronger than standard network nonlocality, we prove that the most well-known network Bell test does not witness full network nonlocality. In contrast, we demonstrate that its generalisation to star networks is capable of detecting full network nonlocality in quantum theory. More generally, we point out that established methods for analysing local and theory-independent correlations in networks can be combined in order to systematically deduce sufficient conditions for full network nonlocality in any network and input/output scenario. We demonstrate the usefulness of these methods by constructing polynomial witnesses of full network nonlocality for the bilocal scenario. Then, we show that these inequalities can be violated via quantum Elegant Joint Measurements.
\end{abstract}

\maketitle

%%%%%%%%%%%%%%%%%%%%%%%%%%%%%%%%%%%%%%%%%%%%%%%%%%%%%%%%%%%%%%%%%%%

A network connects a number of parties using some configuration of independent sources. They are common in standard information technology (e.g.~the internet) and their quantum counterparts have been developing rapidly in recent years \cite{Wehner2018, Liao2018, Yu2020}. In a quantum network, a source may distribute entanglement to a set of parties, who can then use entangled measurements to further propagate it along the network \cite{Zukowski1993}. In the last decade, much research in quantum information has been dedicated to understanding correlations in networks (see Ref.~\cite{ReviewPaper} for a review).

\begin{figure}
	\includegraphics[width=0.7\columnwidth]{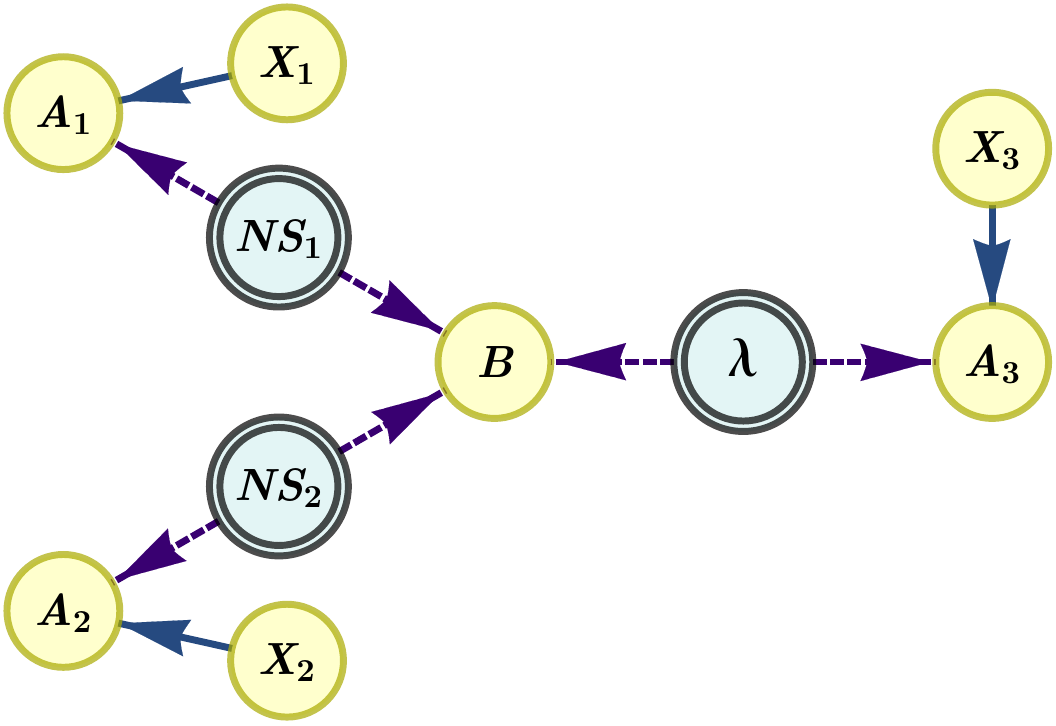}
	\caption{Star network with three branches. Each branch connects a party ($A_1,A_2,A_3$) to the central party ($B$) via an independent source. All parties perform local measurements, for which the branch parties receive inputs ($X_1,X_2,X_3$). The outcome statistics is said to be fully network nonlocal if it cannot be explained in any model of the network where one of the three sources is represented by a classical random variable $\lambda$ and the other two by general no-signaling resources (NS).}
	\label{fig:ThreeStar}
\end{figure}

Let a network be composed of $m$ sources and $n$ parties, each of whom selects a private input $x_k$ and produces an output $a_k$ (see, e.g.,~Figure~\ref{fig:ThreeStar}). The resulting correlations are said to admit a network local model if they can be understood by each source independently emitting a local variable $\lambda_j$:
\begin{equation}\label{Eqlocal}
	\begin{split}
		p(\bar{a}|\bar{x})=&\int d\lambda_1\mu_1(\lambda_1)\ldots \int d\lambda_m \mu_m(\lambda_m) \\
		& \times p(a_1|x_1,\bar{\lambda}_1)\ldots p(a_n|x_n,\bar{\lambda}_n),
	\end{split}
\end{equation}
where $\bar{a}\,{=}\,(a_1,\ldots,a_n)$, $\bar{x}\,{=}\,(x_1,\ldots,x_n)$, $\mu_1,\ldots,\mu_m$ are probability density functions, and $\bar{\lambda}_k$ is the set of local variables associated to the sources that connect to party $k$. This definition reduces to Bell's notion of local causality \cite{Bell1964,Svetlichny1987} when the network is trivial, i.e.~when it only has a single source connecting all parties. If a correlation $p(\bar{a}|\bar{x})$ does not admit a model of the form \eqref{Eqlocal}, it is called network nonlocal (NN).

While it is generally believed that networks enable new forms of nonlocality as compared to standard Bell scenarios (see, e.g.,~\cite{Branciard2010,Fritz2012,Renou2019,ComplexNumber}), it remains largely unclear how such phenomena arise and to what extent they are intrinsic to the network structure. This is partly due to properties inherent to the definition \eqref{Eqlocal}. For instance, if just two parties in a large network violate the celebrated Clauser-Horne-Shimony-Holt (CHSH) Bell inequality \cite{CHSH1969}, the generated distribution is NN regardless of the (perhaps quite trivial) correlations in the rest of the network. Such network nonlocality is arguably neither conceptually novel nor truly a network phenomenon. This motivates the need for conceptualising stronger notions of nonlocality in networks.

Here, we introduce the concept of full network nonlocality. It is defined as follows.
\begin{quote}
	\textit{In a given network and input/output scenario, $p(\bar{a}|\bar{x})$ is fully NN iff it cannot be modelled by allowing at least one source in the network to be of a local-variable nature, while all other sources may be general independent nonlocal resources.}
\end{quote}
We focus here on a literal definition due to the known difficulty of insightfully describing correlations produced by general nonlocal resources in networks \cite{Henson2014,coiteux2021}, and we provide concrete mathematical formulations in specific scenarios throughout the work.

Put in other words, correlations $p(\bar{a}|\bar{x})$ that require all sources to be nonlocal resources are said to be fully NN.
Such sources may emit entangled quantum states, but can also correspond to more general nonlocal resources \cite{Barrett2007}.
Hence, full NN is defined independently of quantum theory.
In the spirit of many works in the device-independent paradigm (see, e.g.,~\cite{Barrett2005,Pironio2010}), this implies that the interest and relevance of quantum demonstrations of full NN do not hinge on the assumption that nature supports no stronger correlations than those of quantum theory.
Full NN is clearly a notion strictly stronger than standard NN.
Furthermore, full NN more faithfully adopts the perspective that nonlocality is a property relative to a network: correlations $p(\bar{a}|\bar{x})$ that are fully NN in a given network are not guaranteed to be if the network is expanded.
This is not the case for standard NN, which is a property that is preserved when the network is enlarged.

This simple concept motivates several elementary questions. How, and to what extent, is full NN different from standard NN? Can we re-examine established network Bell experiments through the lens of full NN? How can we detect full NN, and is a general characterisation possible? Does quantum theory allow for fully NN correlations? If yes, what are the experimental implications? In the following, we address these questions.

\begin{figure}
	\centering
	\includegraphics[width=0.8\columnwidth]{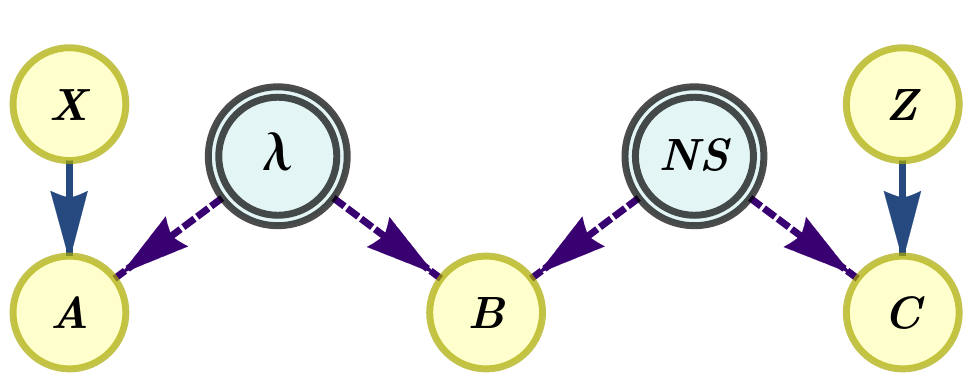}
	\caption{Bilocal scenario in which Alice and Bob share a random variable ($\lambda$) and Bob and Charlie share a general nonlocal resource (NS). Even though there is only one nonlocal resource in the network, correlations in this scenario (see Eq.~\eqref{biloc}) can simulate every possible violation of the standard bilocal Bell inequality.}
	\label{FigBiloc}
\end{figure}

We begin by showing that full NN has major implications for the most well-known network Bell test. Consider the simplest network, known as the bilocal scenario, which features three parties. Alice and Bob are connected via a source and Bob and Charlie via another independent source (see Figure~\ref{FigBiloc}). Alice and Charlie each have binary inputs, $x,z\,{\in}\,\{0,1\}$, and produce binary outcomes, $a,c\,{\in}\,\{0,1\}$, while Bob has a fixed measurement with four possible outcomes, $b\,{\equiv}\, (b_0,b_1)\,{\in}\,\{0,1\}^2$. In the bilocal scenario, models of the form \eqref{Eqlocal} respect the network Bell inequality \cite{Branciard2012}
\begin{equation}\label{brgp}
	\mathcal{S}_2\coloneqq \sqrt{|I_0|}+\sqrt{|I_1|}\leq 1,
\end{equation}
where $I_t\,{\coloneqq}\,\frac{1}{4}\sum_{a,b,c,x,z}(-1)^{a+b_t+c+t(x+z)}p(a,b,c|x,z)$ for $t\,{\in}\,\{0,1\}$. Thus, a violation of Eq.~\eqref{brgp} implies NN. The largest known quantum violation is $\mathcal{S}_2\,{=}\,\sqrt{2}$ and it is obtained by each source emitting the singlet state \mbox{$\ket{\psi^-}\,{=}\,(\ket{01}-\ket{10})/\sqrt{2}$}, Bob projecting onto the Bell basis and Alice and Charlie measuring suitable anticommuting observables \cite{Branciard2012}. Equation \eqref{brgp} has inspired many other network Bell inequalities (see, e.g.,~\cite{Mukherjee2016,Tavakoli2016,Andreoli2017,Gisin2017,Tavakoli2017,Luo2018}), and has been (sometimes up to small modifications) the focus of several experiments \cite{Saunders2017,Carvacho2017,Sun2019}.

We now show that it is possible to reproduce the quantum violation $\mathcal{S}_2\,{=}\,\sqrt{2}$ using only a single nonlocal source. Consider that Bob and Charlie share a Popescu-Rohrlich (PR) box \cite{Popescu1994}. Let Alice and Bob share a binary local variable, $\lambda\,{\in}\,\{0,1\}$ with $p(\lambda)\,{=}\,\frac{1}{2}$, and have Alice determine her output as $a\,{=}\,x\lambda$. Bob uses $\lambda$ as his input for the PR box. Thus $c\oplus b'\,{=}\,z\lambda$, where $b'\,{\in}\,\{0,1\}$ is the output of Bob's part of the PR box. Finally, Bob chooses his output to be $b_0\,{=}\,b_1\,{=}\,b'$.
It is easily verified that, for this strategy, $I_0\,{=}\,I_1\,{=}\,\frac{1}{2}$ and consequently that $\mathcal{S}_2\,{=}\,\sqrt{2}$.

Thus, every known quantum violation of the inequality \eqref{brgp} can be simulated in a bilocal network with one local-variable source, if the other source is a general nonlocal resource. However, correlations arising in this scenario from more general probabilistic theories may still be fully NN. How can we characterise the most general non-full NN correlations in the space of $(I_0,I_1)$? Reference \cite{Branciard2012} showed that standard no-signaling correlations (without requiring independent sources) satisfy $|I_0|+|I_1|\leq 1$. By varying $p(\lambda)$ and considering outputs flipping in our strategy, we can generate every pair $(I_0,I_1)$ that satisfies this inequality. Thus, the inequality serves as a tight constraint on non-full NN correlations.

However, although the projection of the correlations in the $(I_0,I_1)$-plane is a convex set, the set of non-full NN correlations is in general non-convex. To see this, we can formalise the concept by defining non-full NN correlations as those admitting the model in Figure~\ref{FigBiloc}, namely,
\begin{equation}\label{biloc}
	p(a,b,c|x,z)=\int d\lambda q(\lambda) p(a|x,\lambda)p(b,c|\lambda,z),
\end{equation}
where $q(\lambda)$ is a probability density, $p(a|x,\lambda)$ is a conditional probability distribution and $p(b,c|\lambda,z)$ is a bipartite no-signaling distribution, i.e.~it satisfies \mbox{$\sum_b p(b,c|\lambda,z)=p(c|z)$} and \mbox{$\sum_c p(b,c|\lambda,z)=p(b|\lambda)$}. In general, to establish full NN, one needs also to consider the scenario when the sources are interchanged, since full NN requires that no simulation of the correlations is possible with a local-variable source anywhere in the network. Equation \eqref{biloc} is straightforwardly extended also to other networks [see, e.g.,~Eq.~\eqref{3starmodel} for a model for the network in Figure~\ref{fig:ThreeStar}]. The non-convexity of the set of correlations follows from the independence of the sources. A simple example is that the distributions $p_k(a,b,c)=\delta_{a,k}\delta_{b,k}\delta_{c,k}$ for $k\,{\in}\,\{0,1\}$ are both compatible with Eq.~\eqref{biloc} but their uniform mixture is fully NN.

Naturally, however, the pivotal question is whether there exist full NN witnesses that can be violated in quantum theory. Here we consider two approaches: one based on entirely different quantum protocols in the bilocal scenario, and another that extends the ideas behind the inequality \eqref{brgp} to larger networks. We begin with the latter, considering star-shaped networks (see Figure~\ref{fig:ThreeStar}). Star networks are composed of $n$ branch parties, which we consider to have binary inputs $(x_1,\ldots,x_n)$ and binary outputs $(a_1,\ldots,a_n)$. The branch parties are individually linked to a single party who performs a fixed measurement with an outcome represented by an $n$-bit string $b\,{\equiv}\, (b_1,\ldots ,b_n)\,{\in}\,\{0,1\}^n$. Generalising the inequality \eqref{brgp}, every model of the form \eqref{Eqlocal} respects the following network Bell inequality \cite{Tavakoli2014}:
\begin{equation}\label{tsca}
	\mathcal{S}_n\coloneqq \frac{1}{2^{n-2}}\sum_{t=1}^{2^{n-1}} |I_t|^{1/n}\leq 1,
\end{equation}
where $I_t\,{\coloneqq}\,\frac{1}{2^n}\sum_{\bar{a},\bar{x},b}(-1)^{a_1+\ldots+a_n+\tilde{t}\cdot b+\hat{t}\cdot \bar{x}}p(\bar{a},b|\bar{x})$, the $n$-bit strings $\tilde{t}$ and $\hat{t}$ are \mbox{$\tilde{t}=(1,t_1,t_2,\dots)$} and \mbox{$\hat{t}=\left(\bigoplus_{k=1}^{n-1} t_k,t_1,t_2,\dots\right)$}, and $(t_1,t_2,\dots)$ is the string of \mbox{$n\,{-}\,1$} bits representing $t$. The best known quantum protocol achieves $\mathcal{S}_n\,{=}\,\sqrt{2}$ by having each source emitting the singlet state $\ket{\psi^-}$, the central party performing a joint $n$-qubit entanglement swapping measurement, and letting the branch parties measure suitable anticommuting observables \cite{Tavakoli2014}. The inequality (or modifications of it) has been violated in experiments using complete entanglement swapping \cite{Baumer2020} and separable measurements \cite{Poderini2020experimental}.

We focus on $n\,{=}\,3$. In Appendix~\ref{App3star} we show that $|I_1|+|I_2|+|I_3|+|I_4|\leq 1$ is a tight constraint satisfied by all non-full NN correlations. In particular, it implies the full NN witness:
\begin{equation}\label{tscaNN}
	\mathcal{S}_3\leq 2^{1/3},
\end{equation}
meaning that a violation implies full NN. Importantly, it follows that quantum correlations reaching $\mathcal{S}_3\,{\in}\, (1,2^{1/3}]$ are certified to be NN but not fully NN, whereas quantum correlations reaching $\mathcal{S}_3\,{\in}\, (2^{1/3},\sqrt{2}]$ are both NN and fully NN. Because of the sizable gap between the bound on the right-hand side of Eq.~\eqref{tscaNN} and the largest known quantum value of $\mathcal{S}_3$, a reasonable degree of noise can also be tolerated in the quantum realisation. If each source emits a Werner state \cite{Werner1989} $\rho_v\,{=}\,v\psi^- +\frac{1-v}{4}\openone$, where $\psi^-\,{=}\,\ketbra{\psi^-}{\psi^-}$ and $v\,{\in}\,[0,1]$ is the visibility, then a quantum demonstration of full NN is possible whenever $v\,{>}\,2^{-1/6}\,{\approx}\, 89.1\%$.

Furthermore, star networks allow us to explore the relationship between standard NN and full NN in a scalable manner, i.e.~by considering arbitrarily large values of $n$. It turns out that the network Bell inequalities \eqref{tsca} reveal differences between the two concepts in a rather extreme manner. To showcase this, imagine that in spite of having access to $n$ sources, only one of them is of a nonlocal nature and the remaining $n\,{-}\,1$ sources correspond to local variables. This is sufficient to violate Eq.~\eqref{tsca} for any $n$. Let each of the parties $k\,{=}\,2,\ldots,n$ be connected to the central party via a local variable $\lambda_k\in\{0,1\}$ [with $p(\lambda_k)\,{=}\,\frac{1}{2}$] and locally assign their output $a_k\,{=}\,x_k\lambda_k$. The central party uses $\lambda_2\oplus\ldots\oplus \lambda_n$ as a binary input for a PR box shared with the remaining branch party; thus yielding $a_1\,{\oplus}\, b'\,{=}\,x_1(\lambda_2\oplus\ldots\oplus \lambda_n)$. The output $b'$ is then used to choose the final output of the central party as $b_1\,{=}\,b'$ and $b_2\,{=}\,\ldots\,{=}\,b_n\,{=}\,0$. This gives $I_t\,{=}\,\frac{1}{2^{n-1}}$ and $\mathcal{S}_n\,{=}\,2^{1/n}$, giving a violation of Eq.~\eqref{tsca} for every $n$. Notably, for $n\,{=}\,3$ it saturates the bound in Eq.~\eqref{tscaNN}. We note this may be surprising since only one nonlocal source is used, instead of two. However, based on some exploration of $\mathcal{S}_4$, we do not believe that correlations outperforming $\mathcal{S}_n\,{=}\,2^{1/n}$ imply full NN when $n\,{>}\,3$. Notably, extending the derivation in Appendix~\ref{App3star} to the case of $n\,{=}\,4$ gives the (perhaps not tight) full NN witness $\mathcal{S}_4\,{\leq}\, \sqrt{2}$, which matches the best known quantum violation.

The fact that quantum full NN is possible through a violation of Eq.~\eqref{tscaNN} in the three-branch star network motivates us to ask whether it can arise already in the simpler bilocal scenario. Because of our previous discussion, a positive answer would need an approach that is different from the one based on Eq.~\eqref{brgp}. Therefore, we consider the quantum correlations reported in Ref.~\cite{Tavakoli2020}. These correlations arise in the bilocal scenario when both sources emit singlet states, Alice and Charlie each perform the three Pauli measurements, and Bob performs an Elegant Joint Measurement. The latter is a projection of two qubits in a partially entangled basis with the defining property that the marginal states for each qubit forms a regular tetrahedron inside the Bloch sphere \cite{Tavakoli2020}. The tetrahedron radius corresponds with a parameter $\theta\,{\in}\,[0,\frac{\pi}{2}]$. Notably, $\theta\,{=}\,0$ corresponds to the original Elegant Joint Measurement introduced in Ref.~\cite{Gisin2019} and $\theta\,{=}\,\frac{\pi}{2}$ corresponds to the Bell state measurement. In Appendix~\ref{AppEJM} we detail this measurement family and the corresponding quantum correlations, which we denote $p_\theta(a,b,c|x,z)$.

In order to investigate whether $p_\theta(a,b,c|x,z)$ is fully NN for some $\theta$, we adopt a method based on inflation of networks \cite{wolfe2019inflation}. Inflation is a technique by which the sources and measurement devices of a network are copied several times and arranged in different configurations. By studying the correlations that arise in these configurations, one can impose nontrivial constraints on the correlations that can be generated in the original network. Note that inflation is only a theoretical tool and does not mean that the actual network of interest is changed. By considering increasingly large inflations, one obtains increasingly strong conditions for the correlations of interest. Concretely, inflation-based methods have been developed to characterise correlations in networks where all sources are local \cite{wolfe2019inflation}, quantum \cite{wolfe2019qinflation}, and general nonlocal resources \cite{wolfe2019inflation,Gisin2020}. By combining the ideas from the first and last, one obtains a computationally viable and generally applicable tool to analyse correlations in networks that are composed of both local and general nonlocal resources.

We consider the inflation of the bilocal network where the source between Alice and Bob is local and the source between Bob and Charlie is nonlocal. Our inflation is illustrated in Figure~\ref{FigBilocInf} in Appendix~\ref{AppBilocInflation}: following inflations for general nonlocal resources \cite{wolfe2019inflation, Gisin2020} we have duplicated Bob and Charlie and the source connecting them, and following local-variable inflations \cite{wolfe2019inflation} we have copied the local variable originally shared between Alice and Bob. Via standard procedure \cite{wolfe2019inflation, Gisin2020}, sufficient conditions for the incompatibility of $p_\theta$ with the hypothesised model in the bilocal network are obtained through a linear program (see Appendix~\ref{AppBilocInflation} and the computational appendix~\cite{compAppendix}). In this manner, we find that, with the exception of the extreme points $\theta\,{=}\,0$ and $\theta\,{=}\,\frac{\pi}{2}$, the family of correlations $p_\theta$ is fully NN. We note that our inability to detect full NN for $\theta\,{\in}\,\{0,\frac{\pi}{2}\}$ is due to the correlations $p_\theta$ actually not being fully NN. In Appendix~\ref{AppSimulation} we present explicit simulation models for both $p_0$ and $p_{\frac{\pi}{2}}$. That only the less symmetric distributions (in terms of $\theta$) are fully NN is reminiscent of the fact that less entangled bipartite states are harder to simulate, e.g.~with limited communication, with one PR box, exploiting the detection loophole or the EPR2 decomposition \cite{Brunner2005, Methot2007}.

By considering the duality theory of linear programming (see Appendix~\ref{AppWitnesses}), we can convert our proofs of full NN for $p_\theta$ into witnesses of full NN that apply not only to $p_\theta$ but to general distributions in the given network and input/output scenario. Although we have obtained many different witnesses in this way, we highlight one that is particularly elegant. A simultaneous violation of both the following inequalities implies full NN in the bilocal scenario where Alice and Charlie perform three dichotomic measurements and Bob performs a single measurement:
\begin{multline} \label{EJM1}
	-\expect{A_1B_2C_3}-\expect{A_2B_2}\\
	+\expect{C_3}\left[\expect{A_1B_2}+\expect{A_2B_2C_3}+\expect{C_3}\right]\leq 1
\end{multline}
and
\begin{multline} \label{EJM2}
	-\expect{A_1B_2C_3}+\expect{B_2C_2}\\
	+\expect{A_1}\left[\expect{B_2C_3}-\expect{A_1B_2C_2}+\expect{A_1}\right]\leq 1,
\end{multline}
see the computational appendix~\cite{compAppendix} for their derivation. Generally, one may expect one network Bell inequality per arrangement of the local-variable source. A violation of Eq.~\eqref{EJM1} implies that the correlation does not admit a model of the type \eqref{biloc}, and a simultaneous violation of Eq.~\eqref{EJM2} implies that the correlation neither can be generated if the source between Alice and Bob is nonlocal and the source between Bob and Charlie is local. The three-party expectation values are defined as $\expect{A_xB_yC_z}\,{=}\,\sum_{a,b,c}ab_yc\, p(a,b,c|x,z)$, where $a,c\,{\in}\,\{\pm 1\}$ and $b_y$ is the $y$th element in the bit string $(b_1,b_2,b_3)\,{\in}\,\{\pm 1\}^3$ that satisfies $b_1b_2b_3\,{=}\,1$ (this is a handy way to represent Bob's outputs). The two-party and one-party expectation values are defined analogously. However, notice that the inequalities effectively only involve two outcomes on Bob.

We exemplify the use of the inequalities (\ref{EJM1}-\ref{EJM2}) by detecting full NN in the quantum correlations based on Elegant Joint Measurements. To also take noise into the analysis, we consider that the quantum protocol is performed with each source emitting a Werner state $\rho_v$. For simplicity, we let both Werner states have the same visibility. A direct calculation (see Appendix~\ref{AppEJM}) gives that both the left-hand sides in Eqs.~(\ref{EJM1}-\ref{EJM2}) are equal to $\frac{1}{2}v\left(v+v\sin \theta+\cos\theta\right)$. The critical visibility per source required for a violation of Eqs.~(\ref{EJM1}-\ref{EJM2}) is, thus,
\begin{equation}
	v_\text{crit}=\frac{4}{\cos\theta+\sqrt{8+8\sin\theta +\cos^2\theta}}.
\end{equation}
We have $v_\text{crit}\,{<}\,1$ for every $\theta\,{\in}\, (0,\frac{\pi}{2})$. Moreover, the best tolerance to noise is obtained at $\theta\,{=}\,\arccos\left(\frac{\sqrt{5}}{3}\right)\,{\approx}\, 0.7297$ for which it becomes $v_\text{crit}\,{=}\,\frac{2}{\sqrt{5}}\,{\approx}\, 0.8944$. Conversely, if we are given sources of visibility $v$, the best choice of measurement is given by $\theta\,{=}\,\arctan\left(v\right)$. We have also  conducted a numerical search for a simultaneous quantum violation of the inequalities (\ref{EJM1}-\ref{EJM2}) and found no improvement on the presented protocol. Even though the critical visibility here is slightly larger than that required for violating \eqref{tscaNN}, it is important to note that this scenario is considerably simpler: it requires fewer copies, fewer parties and a simpler joint measurement. Also, it highlights the relevance of more general entangled measurements in network nonlocality. Finally, note that the best known visibility for standard NN in the bilocal scenario is $v=\frac{1}{\sqrt{2}}$ \cite{Branciard2012}.

We now discuss some practical and conceptual aspects of full NN, as well as future challenges. An interesting future task is to experimentally demonstrate full NN. To this end, one may consider the violation of one of our above discussed full NN witnesses. In fact, the Elegant Joint Measurement has been implemented in both optics and superconducting circuits \cite{Baumer2020, Tang2020}. Nevertheless, our witnesses are tailored to appealing theoretical properties, rather than experimental friendliness. Therefore, we present in Appendix~\ref{AppExperiment} a quantum protocol for full NN in the bilocal scenario that can be implemented in photonic systems using only linear optics and no auxiliary photons. Since our protocol requires only pairs of entangled qubits, a partial Bell state measurement and a visibility per source of $v\gtrsim 0.92$, we believe a proof-of-principle violation is well within reach of present technology. We note that previous entanglement-swapping-based demonstrations of NN do not constitute demonstrations of full NN; either due to the fundamental reasons already discussed or due to insufficient magnitudes of violations (see, e.g.,~\cite{Baumer2020}). We mention also that the traditional entanglement swapping protocol \cite{Zukowski1993} combined with an event-ready violation of the CHSH inequality \cite{Jennewein2001} (which significantly pre-date the modern topic of Bell nonlocality in networks) may be interpreted as a proof of both standard NN and full NN in the bilocal scenario. However, alike most other works on this topic, our interest is to consider qualitatively different quantum protocols. Nonetheless, it remains an interesting open problem to pinpoint a natural notion of nonlocality in networks that discards event-ready protocols.

An interesting extension of our work is to consider full nonlocality in quantum networks. This amounts to assuming that all sources in a network are described by quantum theory and attempt to identify correlations that necessitate entanglement in all sources. Although this notion is arguably less fundamental, it may be a relevant consideration for quantum information technologies. In analogy with our approach here, we expect that such correlations can be characterised by combining the inflation techniques reported in Refs.~\cite{wolfe2019inflation,wolfe2019qinflation}.

Full network nonlocality is rightfully viewed as a first step in a quest to identify notions of network nonlocality that are more ``genuine'' than the standard definition \eqref{Eqlocal}. It is not unreasonable to suspect that many conceptually different forms of network nonlocality are yet to be identified, with features desirable in different situations. For instance, full NN is not stable under composition, so distributions that are not fully NN can become so by grouping parties. In this aspect, full NN is not different from genuinely multipartite nonlocality~\cite{Svetlichny1987} or genuinely multipartite entanglement~\cite{Seevinck2001,Navascues2020}. A particularly interesting next step would be to consider correlations that not only necessitate all sources to be nonlocal, but also assert the role of entanglement swapping on the level of the correlations.

\begin{acknowledgments}
We are grateful to Cyril Branciard for very insightful comments.
A.P.-K.~is supported by the European Union's Horizon 2020 research and innovation programme-grant agreement No. 648913 and by the Spanish Ministry of Science and Innovation through the ``Severo Ochoa Programme for Centres of Excellence in R\&D'' (CEX2019-000904-S).
N.G.~is supported by the Swiss National Science Foundation via the National Centres of Competence in Research (NCCR)-SwissMap.
A.T.~is supported by the Swiss National Science Foundation through Early PostDoc Mobility fellowship P2GEP2 194800 and acknowledges funding from the Wenner-Gren Foundations.
\end{acknowledgments}

\bibliographystyle{apsrev4-2-custom}
\bibliography{fullNN_references}

%%%%%%%%%%%%%%%%%%%%%%%%%%%%%%%%%%%%%%%%%%%%%%%%%%%%%%%%%%%%%%%%%%%
\onecolumngrid
\pagebreak
\appendix

\section{Witnessing full NN in the 3-star network}\label{App3star}
We prove a condition respected by all non-full NN correlations in the space of $(I_1,I_2,I_3,I_4)$ for the star network with three sources. Due to symmetry in the quantities $I_t$, it is sufficient to consider only one of the three configurations of placing one local-variable source in the network. Therefore, we investigate correlations obtained from the following model:
\begin{equation}\label{3starmodel}
	p(a_1,a_2,a_3,b|x_1,x_2,x_3)=\int d\lambda q(\lambda)p(a_1,a_2,b|x_1,x_2,\lambda)p(a_3|x_3,\lambda),
\end{equation}
where $q(\lambda)$ is a probability density function, $p(a_3|x_3,\lambda)$ is a stochastic local response function and $p(a_1,a_2,b|x_1,x_2,\lambda)$ is a three-partite theory-independent distribution in the subnetwork corresponding to the two relevant branch parties and the central party. This distribution must obey the no-signaling principle as well as the the independence condition between the two involved branch parties, namely $\sum_b p(a_1,a_2,b|x_1,x_2,\lambda)=p(a_1|x_1)p(a_2|x_2)$.

Evaluating the quantities $I_t$ in the model \eqref{3starmodel}, we obtain
\begin{align}
	& I_1=\frac{1}{8}\int d\lambda q(\lambda) \sum_{x_1,x_2} \expect{B_1A^1_{x_1}A^2_{x_2}}_\lambda \left(\expect{A_0^3}_\lambda+\expect{A_1^3}_\lambda\right),\\
	&  I_2=\frac{1}{8}\int d\lambda q(\lambda) \sum_{x_1,x_2} (-1)^{x_1}\expect{B_2A^1_{x_1}A^2_{x_2}}_\lambda \left(\expect{A_0^3}_\lambda-\expect{A_1^3}_\lambda\right),\\
	& I_3=\frac{1}{8}\int d\lambda q(\lambda) \sum_{x_1,x_2} (-1)^{x_2}\expect{B_3A^1_{x_1}A^2_{x_2}}_\lambda \left(\expect{A_0^3}_\lambda-\expect{A_1^3}_\lambda\right),\\
	& I_4=\frac{1}{8}\int d\lambda q(\lambda) \sum_{x_1,x_2} (-1)^{x_1+x_2}\expect{B_4A^1_{x_1}A^2_{x_2}}_\lambda \left(\expect{A_0^3}_\lambda+\expect{A_1^3}_\lambda\right),
\end{align}
where the effective correlators appearing are defined by $\expect{B_tA^1_{x_1}A^2_{x_2}}_\lambda=\sum_{a_1,a_2,b}(-1)^{a_1+a_2+\tilde{t}\cdot b}p(a_1,a_2,b|x_1,x_2,\lambda)$ and $\expect{A^3_{x_3}}_\lambda=\sum_{a_3}(-1)^{a_3}p(a_3|x_3,\lambda)$. Evaluating the absolute value, we obtain the bounds
\begin{align}
	& |I_1|\leq \frac{1}{8}\int d\lambda q(\lambda) \left|\sum_{x_1,x_2} \expect{B_1A^1_{x_1}A^2_{x_2}}_\lambda\right| \left|\expect{A_0^3}_\lambda+\expect{A_1^3}_\lambda\right|,\\
	&  |I_2|\leq \frac{1}{8}\int d\lambda q(\lambda) \left|\sum_{x_1,x_2} (-1)^{x_1}\expect{B_2A^1_{x_1}A^2_{x_2}}_\lambda \right| \left|\expect{A_0^3}_\lambda-\expect{A_1^3}_\lambda\right|,\\
	& |I_3|\leq \frac{1}{8}\int d\lambda q(\lambda) \left|\sum_{x_1,x_2} (-1)^{x_2}\expect{B_3A^1_{x_1}A^2_{x_2}}_\lambda\right| \left|\expect{A_0^3}_\lambda-\expect{A_1^3}_\lambda\right|,\\
	& |I_4|\leq \frac{1}{8}\int d\lambda q(\lambda) \left|\sum_{x_1,x_2} (-1)^{x_1+x_2}\expect{B_4A^1_{x_1}A^2_{x_2}}_\lambda\right| \left|\expect{A_0^3}_\lambda+\expect{A_1^3}_\lambda\right|.
\end{align}
Consider now the quantity $|I_1|+|I_2|+|I_3|+|I_4|$. We can bound it in a non-full NN model as follows:
\begin{align}\notag
|I_1|+|I_2|+|I_3|+|I_4|\leq& \frac{1}{8}\Bigg[\int d\lambda q(\lambda) \left(\left|\sum_{x_1,x_2} \expect{B_1A^1_{x_1}A^2_{x_2}}_\lambda\right|+
\left|\sum_{x_1,x_2} (-1)^{x_1+x_2}\expect{B_4A^1_{x_1}A^2_{x_2}}_\lambda\right|\right) \left|\expect{A_0^3}_\lambda+\expect{A_1^3}_\lambda\right| \\\notag
& +\int d\lambda q(\lambda) \left(\left|\sum_{x_1,x_2}(-1)^{x_1} \expect{B_2A^1_{x_1}A^2_{x_2}}_\lambda\right|+
\left|\sum_{x_1,x_2} (-1)^{x_2}\expect{B_3A^1_{x_1}A^2_{x_2}}_\lambda\right|\right) \left|\expect{A_0^3}_\lambda-\expect{A_1^3}_\lambda\right|
\Bigg]\\
=&\frac{1}{8}\int d\lambda q(\lambda)\left(T_1\left|\expect{A_0^3}_\lambda+\expect{A_1^3}_\lambda\right|+T_2\left|\expect{A_0^3}_\lambda-\expect{A_1^3}_\lambda\right|\right),
\end{align}
where in the last line we have just labelled the two paranthesised expressions by $T_1$ and $T_2$ respectively.

Next, for a fixed $\lambda$, we bound the largest values of $T_1$ and $T_2$. To this end, we consider a relaxation of the network structure where we replace the bilocal network structure connecting parties $A^1$ and $A^2$ separately to party $B$ with a standard Bell scenario, in which all three parties share a common source. Thus, we evaluate the largest values of $T_1$ and $T_2$ in a standard no-signaling model. By considering the four sign combinations associated to the two absolute values appearing in $T_1$ (resp. $T_2$), we can solve the problem by means of standard linear programming (optimising over the no-signaling polytope). We find $T_1\,{\leq}\, 4$ and $T_2\,{\leq}\, 4$. Thus, we obtain
\begin{equation}
	|I_1|+|I_2|+|I_3|+|I_4|\leq \frac{1}{2}\int d\lambda q(\lambda)\left(\left|\expect{A_0^3}_\lambda+\expect{A_1^3}_\lambda\right|+\left|\expect{A_0^3}_\lambda-\expect{A_1^3}_\lambda\right|\right).
\end{equation}
Lastly, we use that for arbitrary $0\leq (r,s)\leq 1$, it holds that $|r+s|+|r-s|\leq 2$. We therefore arrive at
\begin{equation}\label{finres}
	|I_1|+|I_2|+|I_3|+|I_4|\leq 1.
\end{equation}
We can now use the result \eqref{finres} to bound the Bell expression $\mathcal{S}_3$:
\begin{equation}
\mathcal{S}_3:=\frac{1}{2}\left[|I_1|^{1/3}+|I_2|^{1/3}+|I_3|^{1/3}+|I_4|^{1/3}\right]\leq \frac{1}{2}\left[|I_1|^{1/3}+|I_2|^{1/3}+|I_3|^{1/3}+\big|1-|I_1|-|I_2|-|I_3|\big|^{1/3}\right].
\end{equation}
The maximal value of the right-hand-side expression is $2^{1/3}$ when optimised over the relevant constraint; $|I_1|+|I_2|+|I_3|\leq 1$.

Note that, in obtaining the bound, we have relaxed the bilocal structure of the sub-network composed by $A^1$, $A^2$, and $B$. This could, in principle, provide a bound that is higher than what is actually achievable. However, this is not the case: every set $(I_1,I_2,I_3,I_4)$ that satisfies this condition can be realised in a non-full NN model based on two PR boxes and a source of local variables.
To this end, consider that the first two branch parties independently share a PR box with the central party. The third party shares a local variable with the central party (see Figure~\ref{fig:ThreeStar}). The local variable is composed of two bits $\lambda=\lambda_1\lambda_2\in\{0,1\}^2$ with distribution $p_\lambda$. The branch party connected with the local variable chooses the output $a_3=(\lambda_1\oplus \lambda_2)x_3$. The central party then uses $\lambda_1$ (resp. $\lambda_2$) as an input in the PR box shared with the first (resp. second) branch party. This leads to correlations $a_1\oplus b'_1=x_1\lambda_1$ and $a_2\oplus b'_2=x_2\lambda_2$, where $b'_1$ and $b'_2$ are the respective binary outcomes of the PR box of the central party. Then, the central party defines $b_2=b_3=0$ and $b_1=b'_1\oplus b'_2$. Then, it is easily checked that the four different choices of $\lambda$ correspond to tuples $(I_1,I_2,I_3,I_4)\in \{(1,0,0,0),(0,1,0,0),(0,0,1,0),(0,0,0,1)\}$. Thus, by varying $p_\lambda$ (and also considering local bit-flips) every tuple satisfying Eq.~\eqref{finres} can be attained.

\section{Quantum correlations based on Elegant Joint Measurements}\label{AppEJM}
Here, we summarise the quantum correlations arising in the bilocal scenario when Bob performs an Elegant Joint Measurement. These were originally reported in Ref.~\cite{Tavakoli2020}.

Consider a bilocal network where Alice and Charlie have ternary inputs $x,z\in\{1,2,3\}$ and binary outputs $a,c\in\{+1,-1\}$. Bob has a fixed input and four possible outcomes. His outcomes are written as a three-bit string $b\equiv (b_1,b_2,b_3)\in\{+1,-1\}^3$ under the constraint that $b_1b_2b_3\,{=}\,1$. Consider that each source emits an independent Werner state $\rho_v\,{=}\,v \psi^-+\frac{1-v}{4}\openone$ of visibility $v$. For simplicity, we let both Werner states have the same visibility.
Alice and Charlie associate their inputs to the three standard Pauli observables $\sigma_1$, $\sigma_2$ and $\sigma_3$, and Bob performs a so-called Elegant Joint Measurement. The Elegant Joint Measurement was first reported in Ref.~\cite{Gisin2019} but was generalised to a one-parameter measurement family in Ref.~\cite{Tavakoli2020}. The parameter is $\theta\in[0,\frac{\pi}{2}]$. Every choice of $\theta$ corresponds to an entangled two-qubit rank-one projective measurement. The measurement basis vectors are
\begin{align}\notag
	&\ket{\psi_1}=\frac{1}{2}\left(e^{\frac{-i\pi}{4}}\ket{00}-r_+(\theta)\ket{01}-r_-(\theta)\ket{10}+e^{\frac{-3i\pi}{4}}\ket{11}\right),\\\notag
	&\ket{\psi_2}=\frac{1}{2}\left(e^{\frac{i\pi}{4}}\ket{00}+r_-(\theta)\ket{01}+r_+(\theta)\ket{10}+e^{\frac{3i\pi}{4}}\ket{11}\right),\\ \notag
	&\ket{\psi_3}=\frac{1}{2}\left(e^{\frac{-3i\pi}{4}}\ket{00}+r_-(\theta)\ket{01}+r_+(\theta)\ket{10}+e^{\frac{-i\pi}{4}}\ket{11}\right),\\
	&\ket{\psi_4}=\frac{1}{2}\left(e^{\frac{3i\pi}{4}}\ket{00}-r_+(\theta)\ket{01}-r_-(\theta)\ket{10}+e^{\frac{i\pi}{4}}\ket{11}\right),
\end{align}
where $r_\pm(\theta)=\frac{1\pm e^{i\theta}}{\sqrt{2}}$ and $(1,2,3,4)$ correspond to the four bit strings indexing Bob's outcome. Thus, the resulting quantum correlations are
\begin{equation}\label{EJMprob}
	p^v_\theta(a,b,c|x,z)=\Tr\left[\left(\frac{\openone+a \sigma_x}{2}\otimes \ketbra{\psi_b}{\psi_b}\otimes \frac{\openone+c \sigma_z}{2}\right)\left(\rho_v\otimes \rho_v\right) \right]
\end{equation}
This is more conveniently written in terms of expectation values. The distribution $p_\theta(a,b,c|x,z)$ is completely characterised by all one-, two- and three-party expectation values: $\expect{A_x}$, $\expect{B_y}$, $\expect{C_z}$, $\expect{A_xB_y}$, $\expect{B_yC_z}$, $\expect{A_xC_z}$ and $\expect{A_xB_yC_z}$. The singles are defined as $\expect{A_x}=\sum_{a,b,c}a\,p(a,b,c|x,z)$, $\expect{B_y}=\sum_{a,b,c}b_y\,p(a,b,c|x,z)$, $\expect{C_z}=\sum_{a,b,c}c\,p(a,b,c|x,z)$. The doubles are analogously given by $\expect{A_xB_y}=\sum_{a,b,c}ab_y\,p(a,b,c|x,z)$, etc., and the triple is similarly given by $\expect{A_xB_yC_z}=\sum_{a,b,c}ab_yc\,p(a,b,c|x,z)$. Notice that due to the network structure, we have $\expect{A_xC_z}=\expect{A_x}\expect{C_z}$. When inserting Eq.~\eqref{EJMprob}, the expectations are calculated to be
\begin{align}
	&\expect{A_x}=\expect{B_y}=\expect{C_z}=0, \quad \expect{A_xB_y}=-\frac{v}{2}\cos\theta \hspace{1mm}\delta_{x,y}, \quad \expect{B_yC_z}=\frac{v}{2}\cos\theta \hspace{1mm}\delta_{y,z}\\
	& \expect{A_xB_yC_z}=\begin{cases}
		-\frac{v^2}{2}\left(1+\sin\theta\right) & \text{if } xyz\in\{123,231, 312\}\\
		-\frac{v^2}{2}\left(1-\sin\theta\right) & \text{if } xyz\in\{132,213, 321\}\\
		0 & \text{otherwise}
	\end{cases}.
\end{align}

\section{Inflation in hybrid networks}\label{AppBilocInflation}
Inflation, introduced by Wolfe et al. \cite{wolfe2019inflation}, is a powerful concept that enables the analysis of correlations in arbitrary causal structures, therefore including networks.
Its underlying mechanism is proving through contradiction: in the context of networks, if a distribution can be generated between some parties by using certain sources, then one can consider which kinds of distributions one would be able to generate if given access to multiple copies of such parties and sources.
The networks that are generated by arranging those copies are called inflations of the original network.

Inflation is motivated by the analysis of compatibility with causal models admitting classical latent variables.
In the context of networks, these correspond to addressing whether the correlations admit a model of the form of Eq.~\eqref{Eqlocal}.
For doing so, inflation crucially exploits the fact that classical information can be cloned: if two parties are maximally correlated via some source sending the same classical information to both of them, this classical information can be copied and sent to a third party, which is now maximally correlated to the remaining two.
This is not the case in quantum mechanics and more general operational-probabilistic theories~\cite{Barrett2007} so, in principle, inflation may not be capable of constraining correlations generated in such theories.
However, it is possible to devise inflations of a network that do not require of information cloning, and thus are not restricted to constraining correlations arising from classical sources.
These inflations are termed non-fanout inflations in Ref.~\cite{wolfe2019inflation}, and they are employed in practise in Ref.~\cite{Gisin2020} to provide theory-independent necessary conditions for correlations to be compatible with the triangle network.

In this work, we consider inflations where both possibilities are included: whenever we want to restrict a source to be classical, we will consider copies of only some of the subsystems it sends to parties, and whenever we want to allow a source to distribute general resources, we will consider independent copies of it, in the sense that one could imagine buying several of those sources from a manufacturer.

\begin{figure}[h]
	\centering
	\includegraphics[width=0.4\columnwidth]{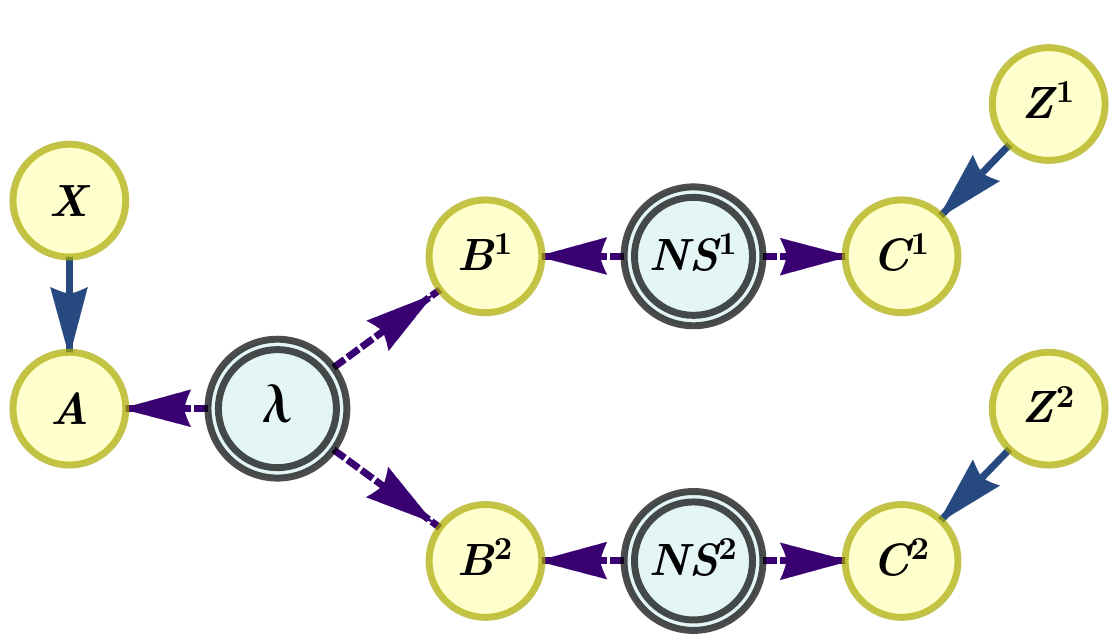}
	\caption{Mixed-resource inflation of the bilocal scenario in Figure~\ref{FigBiloc} when Alice and Bob share a local variable and Bob and Charlie share a general nonlocal resource. The source $\lambda$ is assumed to distribute a classical system, whose subsystems can thus be cloned and sent to copies of Bob. This is not the case for the no-signaling source between Bob and Charlie, for which the no-cloning theorem prohibits the cloning of individual subsystems, but of which we can consider complete copies that connect different copies of Bob and Charlie.}
	\label{FigBilocInf}
\end{figure}

Let us now use inflation arguments to derive necessary conditions for a distribution to admit a model of the type given by Eq.~\eqref{biloc}.
We begin assuming that such model exists, so we know the local variable $\lambda$, characterised by the distribution $q(\lambda)$, the response function of Alice, $p(a|x,\lambda)$, and the no-signaling resource and response functions that produce the distribution $p(b,c|\lambda,z)$, which all combined give rise to $p(a,b,c|x,z)$.
Having access to this information means that we can imagine duplicating the no-signaling resource and Bob's and Charlie's response functions (by going to the manufacturer and buying a new source and measurement devices), and cloning the information from $\lambda$ that is sent to Bob.
Then, we could imagine arranging the available sources and parties in the inflated network depicted in Figure~\ref{FigBilocInf}.
The question is now, what are the properties of the distribution $p_\text{inf}(a,b^1,b^2,c^1,c^2|x,z^1,z^2)$ that is generated in this new network?

For once, the distribution is well defined, i.e.,
\begin{align}
	p_\text{inf}(a,b^1,b^2,c^1,c^2|x,z^1,z^2) &\geq 0 \qquad\forall\,a,b^1,b^2,c^1,c^2,x,z^1,z^2,\label{LPpos}\\
	\sum_{a,b^1,b^2,c^1,c^2}p_\text{inf}(a,b^1,b^2,c^1,c^2|x,z^1,z^2) &= 1 \qquad\forall\,x,z^1,z^2.\label{LPnorm}
\end{align}
Also, the parties are all assumed to be spacelike separated from each other, so no information about any party's input can be received by the rest.
This translates into the constraints
\begin{align}
	\sum_a p_\text{inf}(a,b^1,b^2,c^1,c^2|x,z^1,z^2)-\sum_a p_\text{inf}(a,b^1,b^2,c^1,c^2|x',z^1,z^2)&=0\qquad\forall\,b^1,b^2,c^1,c^2,x,x',z^1,z^2, \notag\\
	\sum_{c^1} p_\text{inf}(a,b^1,b^2,c^1,c^2|x,z,z^2)-\sum_{c^1} p_\text{inf}(a,b^1,b^2,c^1,c^2|x,z',z^2)&=0\qquad\forall\,a,b^1,b^2,c^2,x,z,z',z^2, \label{LPNS}\\
	\sum_{c^2} p_\text{inf}(a,b^1,b^2,c^1,c^2|x,z^1,z)-\sum_{c^2} p_\text{inf}(a,b^1,b^2,c^1,c^2|x,z^1,z')&=0\qquad\forall\,a,b^1,b^2,c^1,x,z^1,z,z'. \notag
\end{align}
Note that here there are no no-signaling constraints involving the copies of Bob because in this particular setup he does not have a choice of measurement to perform, but these constraints can straightforwardly be inserted if necessary.

Moreover, the fact that the sources and parties in the Bob-Charlie sector are indistinguishable copies, and that the exact same information is sent from $\lambda$ to both copies of Bob's measurement device, implies that the distribution is insensitive to the labeling of the sector, this is,
\begin{equation}
	p_\text{inf}(a,b^1,b^2,c^1,c^2|x,z^1,z^2) - p_\text{inf}(a,b^2,b^1,c^2,c^1|x,z^2,z^1) = 0\qquad\forall\,a,b^1,b^2,c^1,c^2,x,z^1,z^2.
	\label{LPinf}
\end{equation}

Finally, note that, if one of the copies of Bob discards its measurement result, we are left in a situation where Alice, the other copy of Bob, and its accompanying Charlie, are exactly reproducing the original network, and there is an additional Charlie that, nevertheless, is still measuring his share of the no-signaling resource.
If the distribution observed in the original network is denoted by $p_\text{orig}(a,b,c|x,z)$, this means that
\begin{equation}
	\sum_{b^2} p_\text{inf}(a,b^1,b^2,c^1,c^2|x,z^1,z^2) = p_\text{orig}(a,b^1,c^1|x,z^1)\,p_\text{orig}(c^2|z^2)\qquad\forall\,a,b^1,c^1,c^2,x,z^1,z^2,\label{LPmarginal}
\end{equation}
where $p_\text{orig}(c|z)\,{=}\,\sum_{a,b}p_\text{orig}(a,b,c|x,z)$ is Charlie's marginal of the distribution observed in the original network.

Recall that the conditions (\ref{LPpos}-\ref{LPmarginal}) are consequences of the correlations admitting a model of the type of Eq.~\eqref{biloc}.
Thus, in order to see whether a distribution $p_\text{obs}$ admits such a model, one can consider the problem
\begin{equation}
	\text{find } p_\text{inf} \text{ such that } \eqref{LPpos},\eqref{LPnorm},\eqref{LPNS},\eqref{LPinf},\eqref{LPmarginal},
	\label{LP}
\end{equation}
which, since all constraints are linear once $p_\text{obs}$ is defined, can be formulated as a linear program.
If a solution to \eqref{LP} does not exist, then it is certified that $p_\text{obs}$ does not admit the conjectured model.
On the contrary, finding a solution may indicate two phenomena: either the constraints derived are too weak to certify that $p_\text{obs}$ does not admit the model, in which case one would resort to consider more copies of the network elements for building larger inflations and to analyse if $p_\text{obs}$ violates any of the constraints implied by those inflations, or $p_\text{obs}$ indeed admits such model, in which case the requirements imposed by all possible valid inflations will be satisfied.

\section{Witnesses from infeasibility certificates}\label{AppWitnesses}
Whenever $p_\text{obs}$ is defined, the problem \eqref{LP} is a linear program.
This means that it can be formulated as $A\cdot\bm{p}_\text{inf}\,{\geq}\,\bm{b}$, where $\bm{p}_\text{inf}$ is a vector containing the variables in the problem, the matrix $A$ contains the coefficients in the left-hand sides of Eqs.~(\ref{LPpos}-\ref{LPmarginal}), and $\bm{b}$ contains the right-hand sides.
Many software packages can be used for calculating a solution, if such solution exists.
Otherwise, Farkas' lemma~\cite{FarkasLemma} guarantees the existence of a vector $\bm{y}$ that satisfies $\bm{y}\cdot A\,{=}\,\bm{0}$ and $\bm{y}\cdot\bm{b}\,{>}\,0$.
Most software packages provide one of such vectors (which is known as a certificate of infeasibility) whenever a solution is identified not to exist.

Interestingly, note that all the dependence of the problem on the observed probability distribution is stored in the vector $\bm{b}$.
This implies that the certificate $\bm{y}$ certifies the infeasibility (which in the problem considered in this work means the existence of full nonlocality) not only of the considered $p_\text{obs}$, but of any probability distribution $p$ that satisfies $\bm{y}\cdot\bm{b}(p)\,{>}\,0$.
By writing the quantity $\bm{y}\cdot\bm{b}(p)$ as a function of the elements of an uncharacterised probability distribution, we obtain the full network Bell inequalities of Eqs.~(\ref{EJM1}-\ref{EJM2}), (\ref{BSM1}-\ref{BSM2}).
In the computational appendix~\cite{compAppendix} we provide computer codes for setting up the problems that produce the certificates which, upon postprocessing, give rise to those inequalities.

\section{Simulation of correlations using one nonlocal source}\label{AppSimulation}
\begin{figure}
	\centering
	\includegraphics[width=0.6\columnwidth]{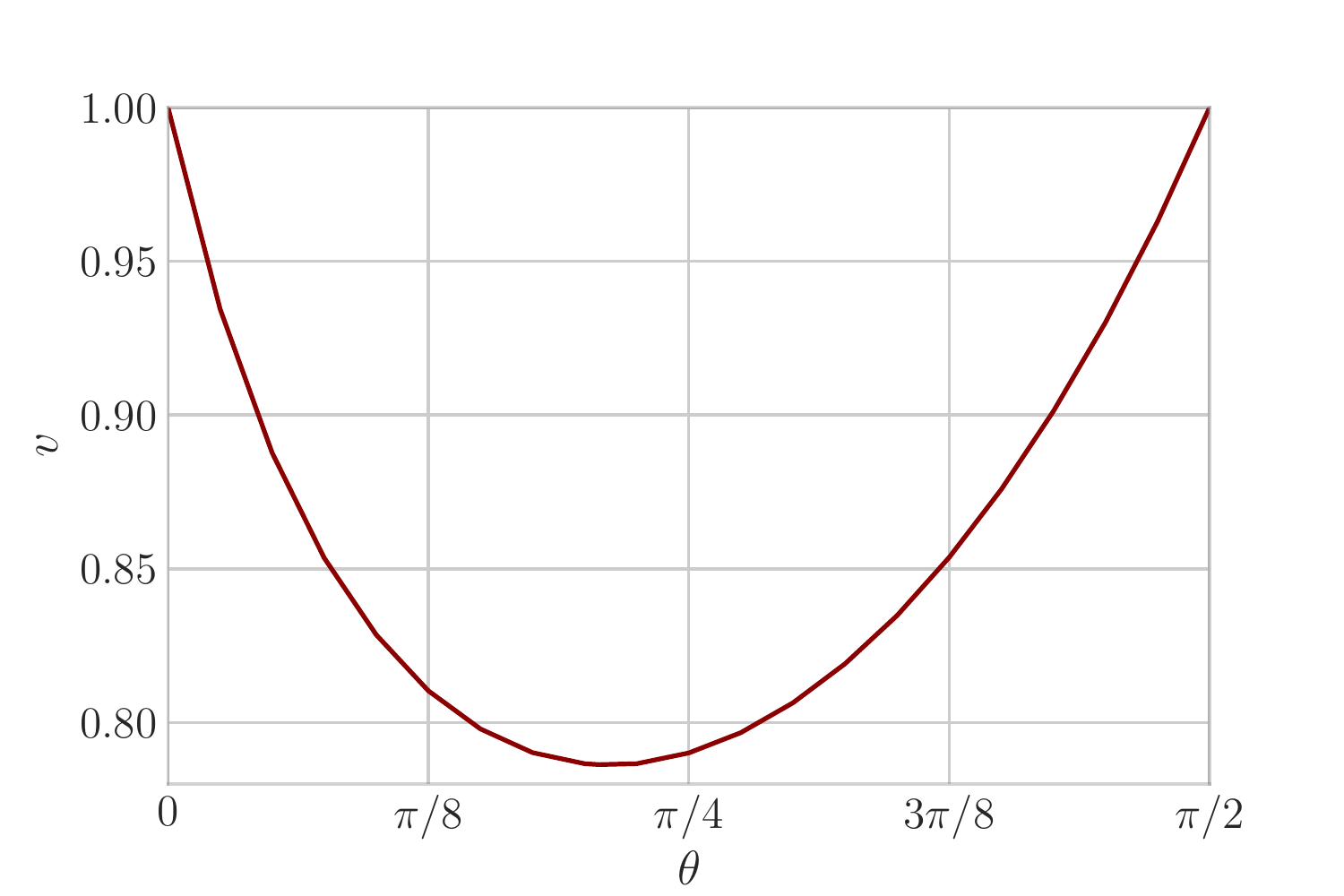}
	\caption{Highest visibility for successful simulation of $p_\theta^v$, given in Eq.~\eqref{EJMprob}, using one local source and one nonlocal source. For the elaboration of this figure, the nonlocal source is placed between Alice and Bob, so the corresponding model that is searched is that given in Eq.~\eqref{biloc}, with the roles of Alice and Charlie interchanged.}
	\label{FigSimulation}
\end{figure}

We have attempted to simulate the quantum correlations $p_\theta^v$ using a model where Alice and Bob share a nonlocal resource and Bob and Charlie share a local variable. Our simulation strategy is as follows. Define
\begin{equation}
m=\begin{pmatrix}
1 & 1 & 1\\
1 & -1 & -1\\
-1 & 1 & -1\\
-1 & -1 & 1
\end{pmatrix},
\end{equation}
which represents vertices of a tetrahedron. Recall that the Elegant Joint Measurements have a tetrahedral symmetry on the collection of their marginal states. We let Bob and Charlie share a four-valued local variable $\lambda\in\{1,2,3,4\}$ with $p(\lambda)\,{=}\,\frac{1}{4}$. We choose Charlie's local response function as $p(c|z,\lambda)\,{=}\,\delta_{c,m_{\lambda,z}}$. Thus, the correlations in the simulation model are written
\begin{equation}
p_\text{sim}(a,b,c|x,z)=\frac{1}{4}\sum_{\lambda=1}^4 \delta_{c,m_{\lambda,z}} p(a,b|x,\lambda),
\end{equation}
where $p(a,b|x,\lambda)$ is a general no-signaling distribution between Alice and Bob. For a given choice of visibility $v$ and measurement $\theta$, we can therefore decide whether a simulation is possible, i.e.~whether $p_\theta^v\,{=}\,p_\text{sim}$ can be satisfied, by solving a linear program over $p(a,b|x,\lambda)$. In this way, we can increase $v$ until the simulation fails.  We have performed this computation for different values of $\theta$ and the results are presented in Figure~\ref{FigSimulation}. The smallest visibility occurs for $\theta\,{\approx}\, 0.65$, for which we find $v\,{\approx}\, 0.7863$. However, there may exist better simulation strategies that can further improve on the visibilities we obtain.

The most interesting results are obtained for $\theta\,{=}\,0$. This corresponds to the case when Bob performs the original Elegant Joint Measurement and $p_0^1$ is known to be network nonlocal \cite{Tavakoli2020}. This distribution is, however, simulated in our model for $v\,{=}\,1$. Therefore, it cannot be fully NN. The simulation is achieved by choosing $p(a,b|x,\lambda)$ as follows.

\begin{equation}
	p(a,b|x,\lambda)=\begin{cases}
		\frac{1}{8} & \text{if } a=m_{\lambda,x}\\
		\frac{1}{2} & \text{if } a\neq m_{\lambda,x} \text{ and } b=\lambda
	\end{cases}.
\end{equation}

Moreover, a complete simulation of $p_\theta^1$ is also possible when $\theta\,{=}\,\frac{\pi}{2}$, which corresponds to Bob implementing a Bell state measurement. The simulation is achieved by choosing $p(a,b|x,\lambda)$ as follows.
\begin{equation}
	p(a,b|x,\lambda)=\begin{cases}
		\frac{1}{4} & \text{if } x=1 \quad \text{and} \quad a\oplus b_1=\lambda_0\oplus\lambda_1\oplus 1\\
		\frac{1}{4} & \text{if } x=2 \quad \text{and} \quad a\oplus b_0\oplus b_1=\lambda_0\oplus 1\\
		\frac{1}{4} & \text{if } x=3 \quad \text{and} \quad a\oplus b_0=\lambda_1\oplus 1\\
		0& \text{otherwise}
	\end{cases},
\end{equation}
where we have written Bob's output as a pair of bits $b_0b_1\in\{00,01,10,11\}$ and similarly written the $\lambda$ as two bits  $\lambda_0\lambda_1\in\{00,01,10,11\}$.

\section{Experimentally friendly witnesses of full network nonlocality}\label{AppExperiment}
We present a full NN witness for the bilocal scenario which admits a quantum violation based on a protocol that is comparatively convenient for implementation in optical systems. We give Alice and Charlie binary inputs and outputs $x,z,a,c\in\{0,1\}$ and let Bob have a fixed input and a ternary output $b\in\{0,1,2\}$. By considering the inflation illustrated in Figure~\ref{FigBilocInf} (and also the case with the sources swapped), we have obtained the following full NN witnesses:
\begin{equation}
	\begin{split}
		\mathcal{R}_\text{C-NS}\coloneqq 2 \langle A_0B_1C_0\rangle -2 \langle A_0B_1C_1\rangle  + 2 \langle A_1B_0C_0\rangle + \langle A_1B_0C_1\rangle - \langle B_0\rangle + \left[\langle A_1B_0\rangle + \langle B_0C_0\rangle - \langle C_0\rangle\right]  \langle C_1\rangle \leq 3
	\end{split}
	\label{BSM1}
\end{equation}
and
\begin{multline}
	\mathcal{R}_\text{NS-C}\coloneqq 2 \langle A_0B_1C_0\rangle - 2\langle A_0B_1C_1\rangle + \langle A_1B_0C_0\rangle + 2\langle A_1B_0C_1\rangle - \langle B_0\rangle \\+ \langle A_1\rangle  \left[\langle A_1B_0\rangle + \langle B_0C_1\rangle + \langle C_0\rangle - \langle C_1\rangle - \langle A_1\rangle\right] \leq 3,
	\label{BSM2}
\end{multline}
where the correlators are computed following Ref.~\cite{Branciard2012}, namely $\langle A_xB_0C_z\rangle\,{=}\,\sum_{a,b,c}(-1)^{a+c}\,[p(a,0,c|x,z)+p(a,1,c|x,z)-p(a,2,c|x,z)]$ and $\langle A_xB_1C_z\rangle\,{=}\,\sum_{a,b,c}(-1)^{a+c}\,\left[p(a,0,c|x,z)-p(a,1,c|x,z)\right]$. These witnesses can be violated with a protocol friendly to linear optics without auxiliary photons. Let both links emit a singlet and let Alice and Charlie measure the observables $A_0\,{=}\,\sigma_x$, $A_1\,{=}\,\sigma_z$, $C_0\,{=}\,\frac{\sigma_z+\sigma_x}{\sqrt{2}}$ and $C_1\,{=}\,\frac{\sigma_z-\sigma_x}{\sqrt{2}}$. Bob performs a partial Bell state measurement defined by $\{\phi^+,\phi^-,\openone-\phi^+-\phi^-\}$, where $\phi^{\pm}=\ketbra{\phi^\pm}{\phi^\pm}$ and $\ket{\phi^\pm}=\frac{\ket{00}\pm\ket{11}}{\sqrt{2}}$. The resulting distribution leads to the violation $\mathcal{R}_\text{C-NS}\,{=}\,\mathcal{R}_\text{NS-C}\,{=}\,5/\sqrt{2}\,{\approx}\, 3.5355$. Moreover, if we substitute the singlets for Werner states $\rho_v$, then a violation is maintained for a visibility per source of $v\,{\geq}\, \sqrt{3\sqrt{2}/5}\,{\approx}\, 0.9212$.

\end{document}